# GeneFEAST: the pivotal, gene-centric step in functional enrichment analysis interpretation

Avigail Taylor[1,2,3*], Valentine M Macaulay[3†], Matthieu J Miossec[2], Anand K Maurya[1] and Francesca M Buffa[4]

[1]Nuffield Department of Medicine, University of Oxford, Old Road Campus, Oxford, OX3 7BN, [2]Centre for Human Genetics, University of Oxford, Roosevelt Drive, Oxford, OX3 7BN, [3]Nuffield Department of Surgical Sciences, University of Oxford, ORCRB, Roosevelt Drive, Oxford, OX3 7DQ, [4]Computational Biology & Integrative Genomics Lab, Department of Oncology, University of Oxford, ORCRB, Oxford, OX3 7DQ

†Deceased.

*To whom correspondence should be addressed.

**Abstract**

**Summary:** GeneFEAST, implemented in Python, is a gene-centric functional enrichment analysis summarisation and visualisation tool that can be applied to large functional enrichment analysis (FEA) results arising from upstream FEA pipelines. It produces a systematic, navigable HTML report, making it easy to identify sets of genes putatively driving multiple enrichments and to explore gene-level quantitative data first used to identify input genes. Further, GeneFEAST can juxtapose FEA results from multiple studies, making it possible to highlight patterns of gene expression amongst genes that are differentially expressed in at least one of multiple conditions, and which give rise to shared enrichments under those conditions. Thus, GeneFEAST offers a novel, effective way to address the complexities of linking up many overlapping FEA results to their underlying genes and data, advancing gene-centric hypotheses, and providing pivotal information for downstream validation experiments.

**Availability and Implementation:** GeneFEAST GitHub repository: https://github.com/avigailtaylor/GeneFEAST; Zenodo record: 10.5281/zenodo.14753734; Python Package Index: https://pypi.org/project/genefeast; Docker container: ghcr.io/avigailtaylor/genefeast.
**Contact:** avigail.taylor@well.ox.ac.uk
**Supplementary Information:** Supplementary information is available at *Bioinformatics* online.

## 1 Introduction

In the era of high-throughput 'omics experiments, functional enrichment analysis (FEA) plays a critical role in our ability to interpret the 'Big' biological data arising from these studies. In a typical workflow, an experiment yields a large set of genes for further analysis (herein, referred to as 'genes of interest', GoI). For example, an RNA-Seq experiment might be used to identify the set of genes differentially expressed between an experimental condition and a control condition. Then, biologically relevant labels are assigned to genes based on some database of terms, pathways or signatures (herein, all referred to as 'terms'), for example, the Gene Ontology (GO) (Ashburner, et al., 2000), or the Kyoto Encyclopedia of Genes and Genomes (KEGG) (Kanehisa and Goto, 2000). Next, FEA is employed to determine which of the biological terms assigned to the GoI are over-represented amongst those genes:

- In an over-representation analysis (ORA) FEA, a hypergeometric test is used to compare the number of GoI annotated by a term to the number of genes annotated by that term amongst the background set of genes assayed in the underlying experiment.

- In a gene set enrichment analysis (GSEA) FEA (Subramanian, et al., 2005, Mootha, et al., 2003), the process is slightly different because the FEA itself helps identify GoI. In particular, all assayed genes are initially considered putative GoI and are ranked by experimental result. For example, in an RNA-Seq experiment genes could be ranked from most over- to most under-expressed in cases versus controls. Then, an enrichment score (ES) is calculated for each term reflecting how often that term's gene set are at the top or bottom of the ranked list. Finally, a *p*-value for the ES is obtained by permutation testing and the 'leading edge' subset of core genes contributing to the term's ES is reported; the superset of all leading edge genes for terms with a significant ES are then the final GoI.

Whichever FEA method is used, in the final step of the FEA workflow, results for all terms are summarised and reports generated (see Supplementary Section 1, Supplementary Figure 1 and Supplementary Table 1 for an overview of the workflow and examples of available tools). This last step is pivotal for researchers to draw biological insights from FEAs, but it is often complicated by the sheer volume of information, which can be multi-dimensional and also contain redundancy.

Importantly, FEAs are usually part of a wider process, contributing to gene-centred hypothesis generation and downstream validation experiments. So, as well as summarising enriched terms, a comprehensive summarisation tool must enable systematic exploration of the link between terms, their associated GoI, and, crucially, gene-level quantitative data first used to identify these genes. Common examples of such data are fold changes or copy number changes in RNA- and DNA-Seq experiments, respectively. Such a tool should highlight gene sets and patterns in quantitative data driving multiple enrichments. Further, it should enable systematic comparison of enrichments found in multiple studies, in terms of patterns in the underlying genes giving rise to these enrichments. Currently, no FEA summarisation tool provides all this functionality (Supplementary Table 2). To fill this gap, we present GeneFEAST: a command-line Python package for summarising and visualising FEA results arising from any standard 'omics database of terms and upstream FEA pipeline.

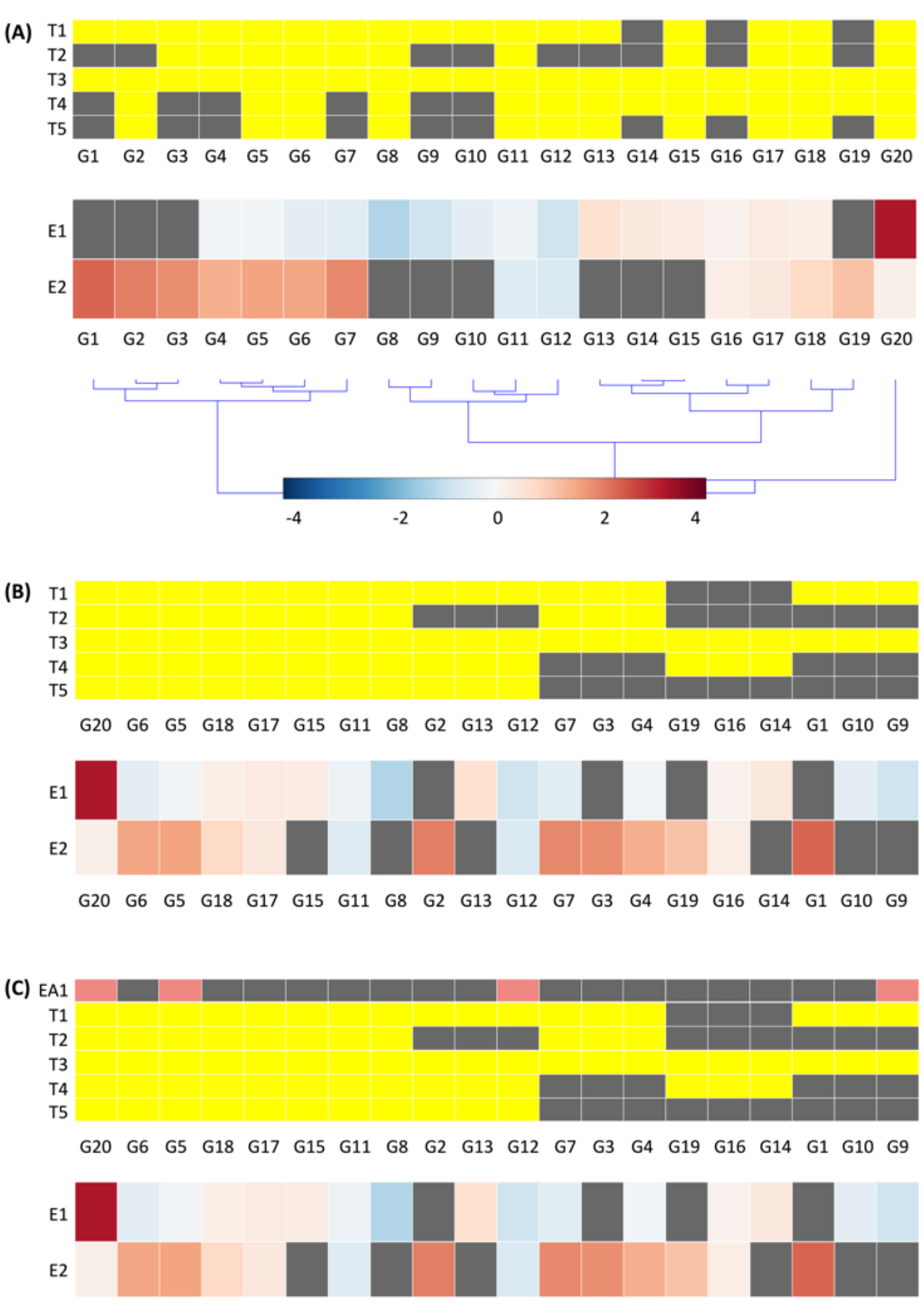

**Fig. 1. Split heatmap. (A)** A pair of heatmaps, sharing a common x-axis of genes, are drawn one on top of the other. In the top heatmap, GoI (G1 to G20) are coloured yellow when they are annotated by a term (T1 to T5), otherwise grey. In the lower heatmap, genes are coloured to depict gene-level quantitative data, in this case log2 fold change from two RNA-Seq experiments whose GoI, i.e., differentially expressed genes, were analysed using ORA-type FEA (E1 and E2). Grey genes were not identified as GoI in the underlying experiment. The genes are ordered based on hierarchical clustering of their quantitative data. **(B)** As for **(A)**, but genes are ordered first by their annotation count, then by annotation pattern, and lastly by their quantitative data. **(C)** A split heatmap with an extra annotation (EA1) added as a row on top of the existing term-GoI heatmap. GoI labelled with the extra annotation are coloured in pink, with the remaining GoI coloured in grey.

## 2 Design and implementation

### 2.1 Grouping terms using gene set overlap

To highlight gene sets driving multiple enrichments, GeneFEAST groups terms into communities using a gene set overlap metric. By using this metric, GeneFEAST remains agnostic to both the 'omics database and upstream FEA used to identify enriched terms.

The grouping algorithm works as follows: First, each term's gene list is reduced to the subset of genes that are GoI's. Next, gene set overlap is calculated between each pair of terms, using either the overlap coefficient (OC) or the Jaccard index (JI), (for two sets $X$ and $Y$, $OC = |X \cap Y|/\min(|X|,|Y|)$, and $JI = |X \cap Y|/|X \cup Y|$), and a network of terms is built with an edge between any pair of terms exceeding a user-defined overlap threshold (Merico, et al., 2010). Within this network, communities of related terms are identified using greedy modularity maximisation (Clauset, et al., 2004), attenuated by an adaptive algorithm that limits the maximum community size (see Supplementary Section 2, Supplementary Figure 2 and Supplementary Boxes 1 to 4 for details). Finally, communities are grouped into larger meta communities when weaker, residual gene set overlap remains between terms from different communities, or when strong gene set overlap exists between terms from different databases, but multi-database agglomeration is off (see Supplementary Box 1 for details). To enable evaluation of community consistency, GeneFEAST outputs a silhouette plot (Rousseeuw, 1987) of communities (Supplementary Figure 3). GeneFEAST also outputs a graphical grid search of community detection parameters to enable a comparison of communities obtained over a range of gene set overlap thresholds and maximum community sizes (Supplementary Figure 4).

Previous approaches have incorporated the idea of clustering terms based on their gene sets to identify broad functional themes in FEAs of one or more experiments (Merico, et al., 2010), or to use this same construction to elucidate complex details of overlapping gene sets giving rise to multiple enrichments (Huang, et al., 2007). The novelty here is in giving the user control over the maximum community size, and in the use of meta communities to address the possibility of terms being placed in multiple communities. Thus, GeneFEAST finds communities of terms that reflect the complex gene set overlaps between terms, ensures that these communities are small enough to be interpretable by the user, and ameliorates loss of information when gene set overlaps span multiple communities.

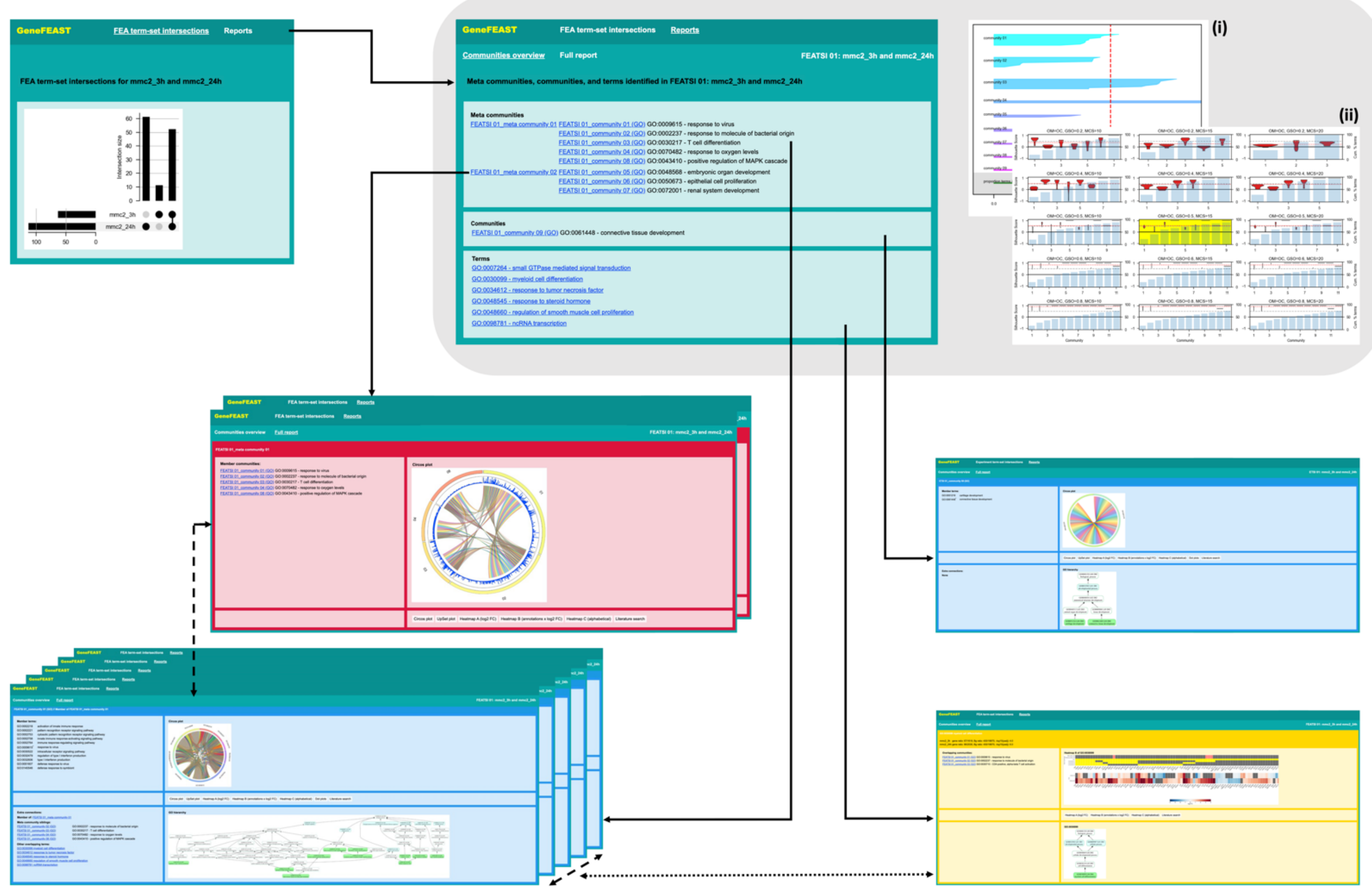

**Fig. 2. Structure and contents of the output HTML/CSS/JavaScript report.** Reports summarising a single FEA have a 'Communities overview' front page (grey inset), which provides a list of meta communities, communities, and terms (green frame in grey inset), a silhouette plot of communities (i), and a graphical grid search of community detection parameters (ii). The Communities overview homepage has anchor links (black, solid arrows) into the 'Full report': there is a link for each meta community (red frames), a separate link to each member community therein (blue frames, bottom left), links to communities of enriched terms that did not form part of a larger meta community (isolated blue frame, right), and links to terms that did not form part of an enriched-term community (yellow frame). A top navigation bar with 'Communities overview' and 'Full report' dropdown menus is fixed at the top of the report and always visible, providing direct access to every part of the report at all times. Reports summarising multiple FEAs start with a front page showing an upset plot of the sets of terms identified as enriched in each of the input FEAs (top left green frame). We refer to each set of terms found in two or more FEAs as a "FEA term-set intersection". The navigation bar at the top of this front page provides a 'Reports' dropdown menu from which the user can navigate to separate reports summarising the terms in each FEA term-set intersection. Each of the separate reports has the structure of a report summarising a single FEA, as described above. Within a GeneFEAST report, every meta community, community and term has a frame of information, implemented in HTML and CSS, which can be scaled to fit the user's monitor. Within each frame, JavaScript enables toggling of content. Meta community frames contain: circos and upset plots showing the gene set overlap of member communities (see User Guide for further details), split heatmaps, wherein term annotation is replaced by gene-community membership in the top heatmap, and a literature search for each gene (as described in the main text). Meta community frames have links to member communities (black, dashed arrow). The content of community frames is described in the main text. Where applicable, community frames have links back to their meta community and also to sibling communities in their meta community (black, dashed arrows); separately, they also have a list of links to terms sharing some gene set overlap, where that overlap was too weak for membership of the community (black, dotted arrow). Term frames have a subset of the content of community frames (see User Guide), and have links back to weakly connected communities (black, dotted arrow).

## 2.2 Split heatmap

GeneFEAST is underpinned by the split heatmap, a data visualisation that we developed (Figure 1). Using this visualisation, we can simultaneously depict term-GoI and experiment-GoI relationships, as well as gene-level quantitative data, for communities of terms and their associated GoI. Crucially, the format can show GoI data from multiple experiments simultaneously, thus enabling a gene-centric comparison of FEA results arising from those multiple experiments. Hierarchical clustering of genes based on their quantitative data highlights global gene-data patterns contributing to enrichments (Figure 1A). Alternatively, ordering genes first by their annotation count, and then by their annotation pattern, highlights subsets of genes contributing to multiple enrichments. Within each of these subsets, genes are then hierarchically clustered based on their quantitative data, thus highlighting local, subset-specific gene-data patterns contributing to enrichments (Figure 1B). Lastly, to enable easy searching of the split heatmap, users can also order genes alphabetically.

### 2.2.1 *A priori* relevant gene sets

Sometimes, users may wish to keep track of an *a priori* set of genes relevant to their study throughout the GeneFEAST report, for example genes contributing to a particular biological signature may be usefully highlighted. In such cases, users can provide extra annotations to be added as rows to the term-GoI heatmap in all split heatmaps of the report (Figure 1C). This is done *post hoc* once communities and meta communities, and their associated GoI, have been identified.

### 2.3 HTML, CSS and JavaScript output report

We required the output report of GeneFEAST to provide intuitive, systematic navigation, visualisation, and review of clustered enriched terms, their associated GoI, and related gene-data and information. To this end, the report is auto-generated in HTML, CSS and JavaScript. Within the report, navigation bars and hyperlinks connect all related information in the document, as well as linking to external websites for further information. Using JavaScript enables a good user experience, because information pertaining to a community, such as figures and tables, can be toggled and viewed in-place, rather than triggering new tabs.

For each community of enriched terms, GeneFEAST reports: (1), member terms; (2), a circos plot (Krzywinski, et al., 2009) and an upset plot (Lex, et al., 2014) showing the overlap between sets of genes annotated by the member terms; (3), split heatmaps of the term- and experiment-GoI relationships, gene-level quantitative data and extra annotations, if supplied; (4), a dot plot summary of member term's FEA results, if supplied; (5), further information about terms, such as automatically generated GO hierarchies and KEGG pathway diagrams, if supplied; (6), external hyperlinks to literature searches for each GoI, via the National Center for Biotechnology Information's Gene and PubMed services (Sayers, et al., 2021) incorporating additional search terms if the user has supplied them, and (7), internal hyperlinks to related communities and terms. (See Figure 2 for further details.)

### 2.4 CSV file output

Term-community membership, term- and experiment-GoI relationships are also output in comma-separated value file format, for input into downstream programs.

## 3 Performance

Table 1 shows typical GeneFEAST runtimes on a bioinformatics-capable laptop, in both single FEA and multiple FEA summarisation modes, when applied to FEA results obtained on publicly available gene expression data (Pinto, et al., 2023).

| GeneFEAST mode | From pip installation | In Docker container |
| --- | --- | --- |
| Single FEA | 3 min 43.95 sec | 3 min 59.23 sec |
| Multiple FEA | 3 min 53.48 sec | 4 min 4.47 sec |

**Table 1. Typical GeneFEAST runtimes**. We timed GeneFEAST (v1.0.0) in single and multiple FEA summarisation modes on a Linux Mint 22 (64-bit) OS laptop, with an 11th Gen Intel i5-1135G7 CPU and 24GB RAM. For the Single FEA time we ran GeneFEAST on 83 GO terms identified as highly significantly enriched (HSE, *p.adjust* < 0.0001) in an ORA of 1780 significantly differentially expressed (DE) genes (*p.adjust* < 0.05) in Calu-3 cells infected with SARS-CoV2 and measured at 3-hours post-infection (hpi) (Pinto, et al., 2023). For the Multiple FEA time we used GeneFEAST to summarise terms common to this first set of 83 GO terms and to a second set of 150 GO terms identified as HSE (*p.adjust* < 0.0001) in an ORA of 2230 significantly DE genes (*p.adjust* < 0.05) in Calu-3 cells also infected with SARS-CoV2, and measured at 24-hpi in the same study. GO ORAs were conducted using the enrichGO function from the clusterProfiler R package (Wu, et al., 2021), with the Benjamini-Hochberg procedure (Benjamini and Hochberg, 1995) applied to control the false discovery rate and adjust *p*-values for multiple testing. For input files and instructions on how to run these performance tests please see Zenodo snapshot 10.5281/zenodo.14773127, directory PERFORMANCE_TEST_INPUT. Note that performance may vary based on hardware and software configuration.

## 4 Running GeneFEAST

GeneFEAST requires Python 3.12 to run and can be pip installed from the Python Package Index at https://pypi.org/project/genefeast. GeneFEAST is also available as a ready-to-use container at ghcr.io/avigailtaylor/genefeast. Viewing the HTML output report requires a web browser with HTML5 and JavaScript 1.6 support. GeneFEAST is OpenSource and available for free; visit http://avigailtaylor.github.io/GeneFEAST for full installation instructions and the User Guide.

## Acknowledgements

We would like to dedicate this paper to our co-author Professor Valentine Macaulay, who sadly passed away before publication of this work. Val was a brilliant scientist and clinician, a fantastic, caring, and supportive boss, and a wonderful friend to us too. She was much loved and is sorely missed.

## Funding

This work has been supported by Cancer Research UK grant C476/A27060 and Prostate Cancer UK grant MA-CT20-006 to VMM, European Research Council microC 772970 grant to FMB, the University of Oxford Returning Carer's Fund to AT and by Wellcome (Core Award 203141/Z/16/Z).

*Conflict of Interest:* none declared.

## References

Ashburner, M., *et al.* Gene ontology: tool for the unification of biology. The Gene Ontology Consortium. *Nat Genet* 2000;25(1):25-29.

Benjamini, Y. and Hochberg, Y. Controlling the false discovery rate: a practical and powerful approach to multiple hypothesis testing. *J R Stat Soc B* 1995;57:289–300.

Clauset, A., Newman, M.E. and Moore, C. Finding community structure in very large networks. *Phys Rev E Stat Nonlin Soft Matter Phys* 2004;70(6 Pt 2):066111.

Huang, D.W., *et al.* The DAVID Gene Functional Classification Tool: a novel biological module-centric algorithm to functionally analyze large gene lists. *Genome Biol* 2007;8(9):R183.

Kanehisa, M. and Goto, S. KEGG: kyoto encyclopedia of genes and genomes. *Nucleic Acids Res* 2000;28(1):27-30.

Krzywinski, M. Circos: An information aesthetic for comparative genomics. *Genome Research* 2009;19:1639–1645.

Lex, A., *et al.* UpSet: Visualisation of Intersecting Sets. *IEEE Trans Vis Comput Graph* 2014;20(12):1983-1992.

Merico, D., *et al.* Enrichment map: a network-based method for gene-set enrichment visualisation and interpretation. *PLoS One* 2010;5(11):e13984.

Mootha, V. K., *et al.* PGC-1alpha-responsive genes involved in oxidative phosphorylation are coordinately downregulated in human diabetes. *Nat Genet* 2003;34(3): 267-273.

Pinto, S.M., *et al*. Multi-OMICs landscape of SARS-CoV-2-induced host responses in human lung epithelial cells. *iScience* 2023;26(1):105895.

Rousseeuw,P.J. Silhouettes: A graphical aid to the interpretation and validation of cluster analysis. *Journal of Computational and Applied Mathematics* 1987;20:53-65.

Sayers, E.W., *et al.* Database resources of the National Center for Biotechnology Information. *Nucleic Acids Res* 2021;49(D1):D10-D17.

Subramanian, A., *et al.* Gene set enrichment analysis: a knowledge-based approach for interpreting genome-wide expression profiles. *Proc Natl Acad Sci U S A* 2005;102(43):15545-15550.

Wu, T., *et al.* clusterProfiler 4.0: A universal enrichment tool for interpreting omics data. *The Innovation* 2021;2(3):100141.